\documentclass[preprint]{revtex4}

\usepackage{graphicx}
\usepackage{dcolumn}
\usepackage{mathrsfs}
\usepackage{slashed}
\usepackage{amsmath,lscape,epsfig}
\usepackage{amsfonts}
\usepackage{amssymb}
\usepackage{graphicx}
\usepackage{dcolumn}
\usepackage{bm}
\def\beq{\begin{equation}}
\def\eeq{\end{equation}}
\def\bea{\begin{eqnarray}}
\def\eea{\end{eqnarray}}
\def\nn{\nonumber}
\def\bge{\begin{equation}}
\def\ene{\end{equation}}
\def\bg{\begin{eqnarray}}
\def\en{\end{eqnarray}}

\def\q{{\bf q}}

\def\q{{\bf q}}
\def\r{{\bf r}}

\def\D0bar{\overline{D^0}}

\begin{document}


\title{$J/\Psi$-nuclear bound states}
\author{
K.~Tsushima$^1$\footnote{kazuo.tsushima@gmail.com},
D.~H.~Lu$^2$\footnote{dhlu@zju.edu.cn,}
G.~Krein$^3$\footnote{gkrein@ift.unesp.br},
A.~W.~Thomas$^1$\footnote{anthony.thomas@adelaide.edu.au}
}
\affiliation{
$^1$ CSSM, School of Chemistry and Physics,
University of Adelaide, Adelaide SA 5005, Australia
\\
$^2$Department of Physics, Zhejiang University, Hangzhou 310027, P.R.China
\\
$^3$Instituto de F\'{\i}sica Te\'orica, Universidade Estadual Paulista \\
Rua Dr. Bento Teobaldo Ferraz, 271 - Bloco II, S\~ao Paulo, SP, Brazil
}

\begin{abstract}
$J/\Psi$-nuclear bound state energies are calculated for a range of
nuclei by solving the Proca (Klein-Gordon) equation.
Critical input for the calculations, namely
the medium-modified $D$ and $D^*$ meson masses, as well as the
nucleon density distributions in nuclei, are obtained
from the quark-meson coupling model.
The attractive potential originates from the $D$ and $D^*$ meson
loops in the $J/\Psi$ self-energy in nuclear medium.
It appears that $J/\Psi$-nuclear bound states should
produce a clear experimental signature
provided that the $J/\Psi$ meson
is produced in recoilless kinematics.
\end{abstract}
\maketitle


\section{Introduction}
\label{sec_intro}

With the $12$~GeV upgrade of the CEBAF accelerator at the Jefferson Lab
in the USA and with the construction of the FAIR facility in Germany,
we expect tremendous progress in understanding
the properties of charmonia and charmed mesons in nuclear medium.
These new facilities will be able to produce low-momenta charmonia
and charmed mesons such as $J/\Psi$, $\psi(2S)$, $D$ and $D^*$
in an atomic nucleus.
Because the targets are nuclei and the nuclear
fermi-momentum is available, it may also be possible to produce these mesons
at subthreshold energies.
While at JLab charmed hadrons will be produced by
scattering electrons off nuclei, at FAIR they will be produced by
the annihilation of antiprotons on nuclei.
One of the major challenges is to find appropriate kinematical conditions
to produce these mesons essentially at rest, or with small momentum
relative to the nucleus. One of the most exciting experimental efforts
may be to search for the $J/\Psi$-nuclear bound states amongst many
other possible interesting experiments in these facilities.
The discovery of such bound states would provide
evidence for a negative mass shift of the $J/\Psi$ meson,
and a possible role of the QCD color van der Waals
forces~\cite{Brodsky:1989jd} in nuclei.
Ref.~\cite{Voloshin:2007dx} presents a recent
review of the properties of charmonium states and compiles a
complete list of references for theoretical studies concerning a
great variety of physics issues related to these states.

There is a relatively long history for the studies of
charmonium binding in nuclei.
The original suggestion~\cite{Brodsky:1989jd} that multiple gluon
QCD van der Waals forces would be capable of binding a charmonium state
led to an estimate of a binding energy as large as $400$~MeV in an $A=9$
nucleus. However, using the same approach but taking into account
the nucleon density distributions in the nucleus,
Ref.~\cite{Wasson:1991fb} found a maximum of $30$~MeV
binding energy in a large nucleus.
Along the same line, including the nuclear
density distributions in nuclei $\eta_c$-nuclear bound state
energies were estimated in Ref.~\cite{deTeramond:1997ny}.
Based on Ref.~\cite{Peskin:1979va},
which showed that the mass shift of charmonium in nuclear matter
can be expressed in terms of the usual
second-order Stark effect arising from the chromo-electric polarizability
of the nucleon, the authors of Ref.~\cite{Luke:1992tm} obtained
a $10$~MeV binding for $J/\Psi$ in nuclear matter,
in the limit of infinitely heavy charm quark mass.
On the other hand, for the excited
charmonium states, a much larger binding energy was obtained, e.g. $700$~MeV
for the $\psi'(2S)$ state, an admittedly untrustworthy
number. Following the same procedure, but keeping the charm quark mass
finite and using realistic charmonium bound-state wave-functions,
Ref.~\cite{Ko:2000jx} found $8$~MeV binding energy for $J/\Psi$ in
nuclear matter, but still over $100$~MeV binding for the charmonium excited
states. While a larger value for the QCD Stark effect is expected for
excited states (because of the larger sizes of these states), the large
values for the binding energies found in the literature for these states
may be an overestimate. A possible source of this overestimate is
the breakdown of the multipole expansion for the
larger-sized charmonium states.

There are some other studies of charmonium interactions with
nuclear matter, in particular involving the $J/\Psi$ meson.
QCD sum rules studies estimated a $J/\Psi$ mass decrease in nuclear matter
ranging from $4$ MeV to $7$ MeV~\cite{Klingl:1998sr,Hayashigaki:1998ey,Kim:2000kj},
while an estimate based on color polarizability~\cite{Sibirtsev:2005ex}
gave larger than $21$ MeV.
Since the $J/\Psi$ and nucleons have no quarks in common,
the quark interchange or the effective meson exchange
potential should be negligible to first order
in elastic scattering~\cite{Brodsky:1989jd},
and multi-gluon exchange should be dominant.
Furthermore, there is no Pauli blocking even at the quark level.
Thus, if the $J/\Psi$-nuclear bound states are formed, the signal for
these states will be sharp and show a clear, narrow peak in the energy dependence
of the cross section, as will be explained later.
This situation has a tremendous advantage compared to the cases of the lighter
vector mesons~\cite{Leupold:2009kz}.

Previously, we have studied~\cite{jpsimass} the $J/\Psi$ mass shift
(scalar potential) in nuclear matter based on an effective Lagrangian
approach, including the $D$ and $D^*$ meson loops in the $J/\Psi$ self-energy.
This is a color-singlet mechanism at the hadronic level,
and may be compared to the color-octet mechanism of
multi-gluon exchange, or QCD van der Waals forces.
In this work, we extend our previous study made in nuclear
matter~\cite{jpsimass} to finite nuclei,
and compute the $J/\Psi$-nuclear bound state
energies by explicitly solving the Proca (Klein-Gordon) equation
for the $J/\Psi$ meson produced nearly at rest.
Furthermore, the structure of heavy nuclei, as well as the
medium modification of the $D$ and $D^*$ masses, are explicitly
included based on the quark-meson coupling
(QMC) model~\cite{Saito:1996sf}.

\begin{figure}[htb]
  \includegraphics[height=0.15\textheight]{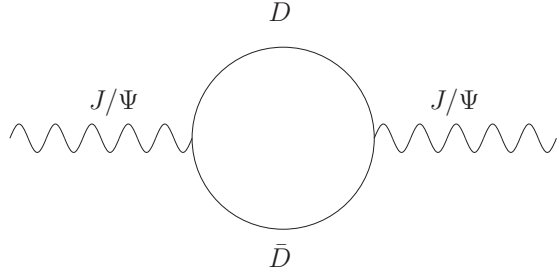}
  \caption{$DD$-loop contribution to the $J/\Psi$ self-energy. We include
  also  $DD^*$ and $D^*D^*$ contributions.}
  \label{fig:loop}
\end{figure}

A first estimate for the mass shifts (scalar potentials) for the
$J/\Psi$ meson (and $\Psi(3686)$ and $\Psi(3770)$) in nuclear medium
including the effects of $D$ meson loop in the $J/\Psi$
self-energy -- see Fig.~\ref{fig:loop} -- was made in Ref.~\cite{Ko:2000jx}.
%
Employing a gauged effective Lagrangian for the coupling of
$D$ mesons to the charmonia, the mass shifts
were found to be positive for $J/\Psi$ and $\psi(3770)$, and negative for
$\psi(3660)$ at normal nuclear matter density $\rho_0$. These results were
obtained for density-dependent $D$ and $\bar D$ masses that decrease linearly
with density, such that at $\rho_0$ they are shifted by $50$~MeV.
The loop integral in the self-energy (Fig.~\ref{fig:loop}) is divergent and
was regularized using form-factors derived from the $^3P_0$ decay
model with quark-model wave functions for $\psi$ and $D$. The positive
mass shift is at first sight puzzling, since even with a $50$~MeV
reduction of the $D$ masses, the intermediate state is still
above threshold for the decay of $J/\Psi$ into a $D\bar D$ pair and
so a second-order contribution should be negative.
However, as we have shown in Ref.~\cite{jpsimass},
this is a result of the interplay of the form factor used and the gauged nature
of the interaction used in Ref.~\cite{Ko:2000jx}.
We have estimated~\cite{jpsimass} the mass shift (scalar potential)
of the $J/\Psi$ meson including the $D\bar D$, $D \bar D^*$, $D^* \bar D$
and $D^*{\bar D}^*$ meson loops in the self-energy using
non-gauged effective Lagrangians. The density dependence
of the $D$ and $D^*$ masses is included by an explicit
calculation~\cite{Tsushima:1998ru} using
the quark-meson coupling (QMC) model~\cite{Guichon:1987jp}.
The QMC model is a quark-based model for nuclear structure which has been very successful
in explaining the origin of many-body or density dependent
effective forces~\cite{qmcskyrme} and hence
describing nuclear matter saturation properties. It has also been used to
predict a great variety of changes of hadron properties in nuclear medium,
including the properties of hypernuclei~\cite{qmchyp-latest}.
A~review of the basic ingredients of the model and a summary of results and
predictions can be found in Ref.~\cite{Saito:2005rv}.

\section{$J/\Psi$ potential in infinite nuclear matter}
\label{sec_potential}

We briefly review how the $J/\Psi$ scalar potential
is calculated in nuclear matter~\cite{jpsimass}.
We use the effective Lagrangian densities at the hadronic level for the vertices,
$J/\Psi{DD}$, $J/\Psi{DD^*}$ and $J/\Psi{D^*D^*}$
(in the following we denote by $\psi$ the field representing $J/\Psi$):
\begin{widetext}
\bea
{\mathcal L}_{\psi D D} &=& i g_{\psi D D} \, \psi^\mu
\left[\bar D \left(\partial_\mu D\right) -
\left(\partial_\mu \bar D\right) D \right] ,
\label{LpsiDDbar} \\
{\cal L}_{\psi D D^*} &=& \frac{ g_{\psi D D^*}}{m_\psi} \,
\varepsilon_{\alpha\beta\mu\nu} \left(\partial^\alpha \psi^\beta\right)
\Bigl[\left(\partial^\mu \bar{D}^{*\nu}\right) D +
\bar D \left(\partial^\mu D^{*\nu}\right)  \Bigr] ,
\label{LpsiDD*} \\
{\cal L}_{\psi D^* D^*} &=& i g_{\psi D^* D^*} \, \bigl\{ \psi^\mu
\left[\left(\partial_\mu \bar{D}^{*\nu}\right) D^*_\nu
- {\bar D}^{*\nu}\left(\partial_\mu D^*_\nu\right) \right]
\nn\\
& &\hspace{3em}+ \left[ \left(\partial_\mu \psi^\nu\right) \bar{D}^*_\nu
- \psi^\nu \left(\partial_\mu {\bar D}^*_\nu\right)
\right] D^{*\mu}
+ \, \bar{D}^{*\mu} \left[\psi^\nu \left(\partial_\mu D^*_\nu\right)
- \left(\partial_\mu \psi^\nu\right) D^*_\nu \right]
\bigr\} .
\label{LpsiD*D*}
\eea
\end{widetext}
The Lagrangian densities above are obtained on the basis of SU(4) invariance
for the couplings among the pseudo-scalar and vector mesons~\cite{Lin,Haidenbauer}.
We use the values for the coupling constants,
$g_{\Psi DD}=g_{\Psi D^*D^*}=g_{\Psi DD^*}=7.64$,
where the former two values were obtained
using the vector meson dominance model~\cite{Lin}.
The form of the $J/\Psi D^*\bar{D}^*$ coupling respects Yang's
theorem~\cite{Yang,Johnson} and so for spin one $J/\Psi$ the virtual
$D^*$ and $\bar{D}^*$ must not be simultaneously transverse.
The difference with the gauged Lagrangian of Ref.~\cite{Ko:2000jx}
for the $\psi D D$ vertex amounts to adding to Eq.~(\ref{LpsiDDbar}) a term
$2 g^2_{\psi D  D} \psi^\mu\psi_\mu \bar D D$.
It may be of interest to investigate in future whether the meson loops
investigated here might play some role in the understanding of the well known
$J/\psi \to \rho \pi$ puzzle~\cite{rhopi_puzzle}.

Only the scalar part contributes to the $J/\Psi$ self-energy,
and the scalar potential for the $J/\Psi$ meson
is the difference of the in-medium, $m^*_\psi$,
and vacuum, $m_\psi$, masses of $J/\Psi$,
\beq
\Delta m = m^*_\psi - m_\psi,
\label{Delta-m}
\eeq
with the masses obtained from
\beq
m^2_\psi = (m^0_\psi)^2 + \Sigma(k^2 = m^2_\psi)\, .
\label{m-psi}
\eeq
Here $m^0_\psi$ is the bare mass and $\Sigma(k^2)$ is the total $J/\Psi$
self-energy obtained from the sum of the contributions from the $DD$, $DD^*$ and
$D^*D^*$ loops. The in-medium mass, $m^*_\psi$, is obtained likewise, with the
self-energy calculated with medium-modified $D$ and $D^*$ meson masses.

\begin{figure}[htb]
\includegraphics[height=95mm,angle=-90]{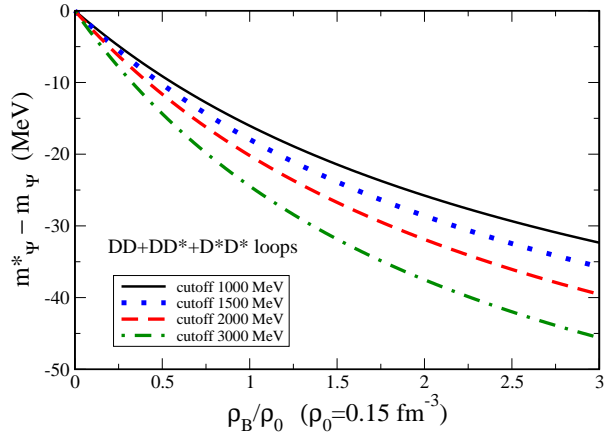}
\caption{The scalar potential arising from $DD$, $DD^*$ and $D^*D^*$ loops
as a function of nuclear matter density for different values
of the cutoff mass $\Lambda_D=\Lambda_{D^*}$~\cite{jpsimass}.}
\label{fig:total}
\end{figure}

In the calculation, we use phenomenological
form factors to regularize the self-energy loop integrals
following a similar calculation for the $\rho$
self-energy~\cite{Leinweber:2001ac}:
\begin{equation}
u_{D,D^*}(\q^2) =  \left(\frac{\Lambda_{D,D^*}^2 + m^2_\psi}
{\Lambda_{D,D^*}^2 + 4\omega^2_{D,D^*}(\q)}\right)^2.
\label{uff}
\end{equation}
For the vertices $\Psi{DD},\Psi{DD^*}$ and $\Psi{D^*D^*}$,
the form factors used are respectively,
$F_{DD}(\q^2)  = u^2_{D}(\q^2)$,
$F_{DD^*}(\q^2) = u_D(\q^2) \, u_{D^*}(\q^2)$, and
$F_{D^*D^*}(\q^2) = u^2_{D^*}(\q^2)$
with $\Lambda_{D}$ and $\Lambda_{D^*}$ being the cutoff masses, and
the common values, $\Lambda_{D}=\Lambda_{D^*}$ are used.
For the calculation in finite nuclei, we chose two values within
the range expected to be reasonable,
$\Lambda_{D}=\Lambda_{D^*}=1500$ and $2000$ MeV to
estimate the model dependence.
The scalar potential, ($m^*_\Psi - m_\Psi$), calculated via Eq.~(\ref{m-psi})
in nuclear matter~\cite{jpsimass}, is shown in Fig.~\ref{fig:total}
as a function of nuclear matter density.

\section{$J/\Psi$ bound states in finite nuclei}
\label{sec_nuclei}

In this section we discuss the case that $J/\psi$ is placed in a finite nucleus.
The nucleon density distributions for
$^{12}$C, $^{16}$O, $^{40}$Ca, $^{90}$Zr and $^{208}$Pb
are obtained using the QMC results
for finite nuclei~\cite{Saito:1996sf}.
For $^4$He, we use the parametrization for the density
distribution obtained in Ref.~\cite{Saito:1997ae}.
Then, using a local density approximation
we calculate the $J/\Psi$ potentials in nuclei.
Note that also the in-medium masses of $D$ and $D^*$
mesons, which are necessary to estimate the $J/\Psi$
self-energy consistently,
are calculated in the QMC model.
We emphasize that the same quark-meson coupling constants are used
between the applied meson mean fields and the light quarks in the nucleon
and those in the $D$ and $D^*$ mesons~\cite{Tsushima:1998ru},
where these coupling constants are calibrated by the nuclear
matter saturation properties~\cite{Saito:2005rv}.

\begin{figure}[htb]
\includegraphics[height=80mm,angle=-90]{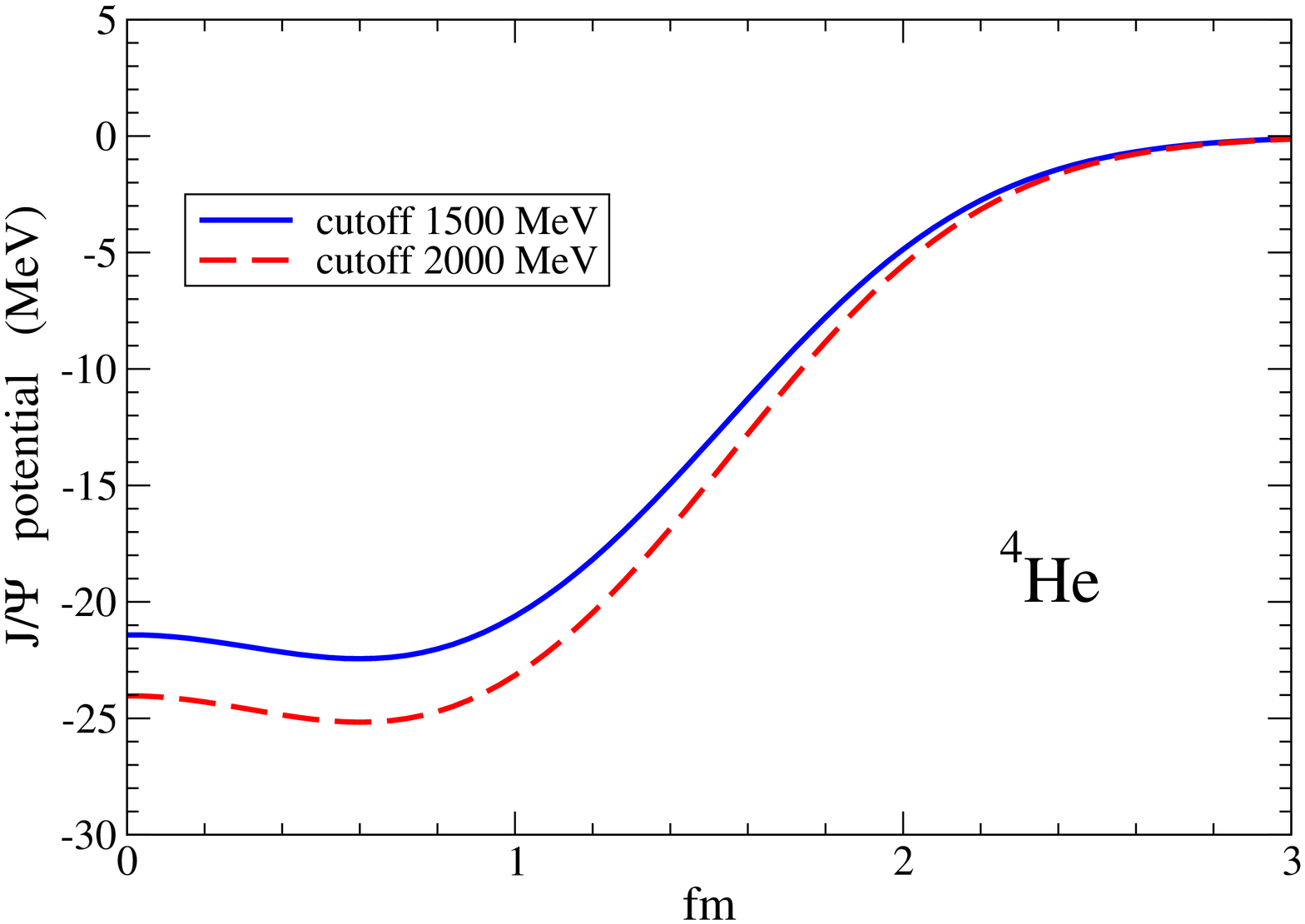}
\includegraphics[height=80mm,angle=-90]{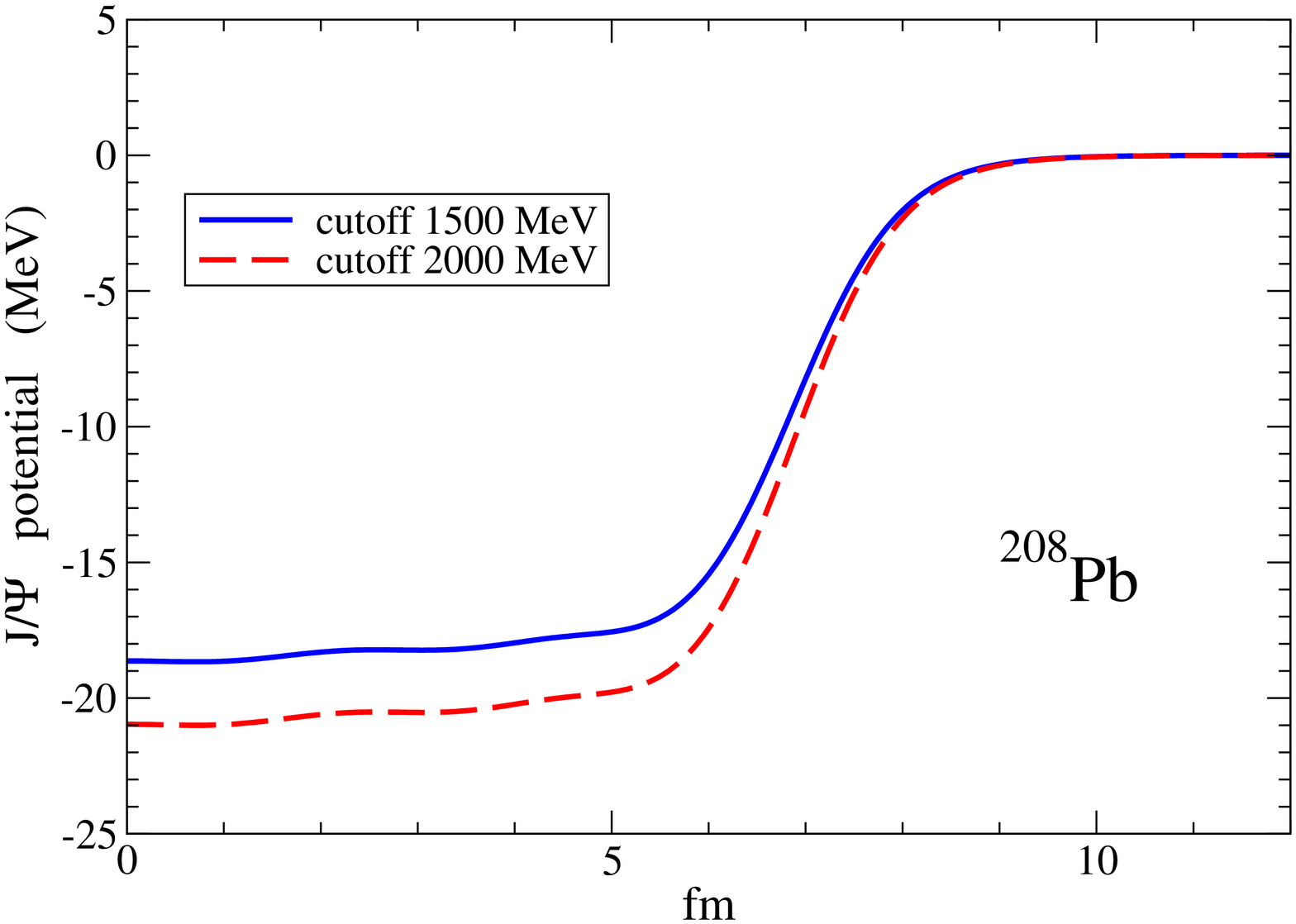}
\caption{
$J/\Psi$ potential in a $^4$He nucleus (left panel)
and in a $^{208}$Pb nucleus (right panel) for two
values of the cutoff mass in the form factors,
$1500$ and $2000$ MeV.
}
\label{fig:poto-nuclei}
\end{figure}

As examples, we show the $J/\Psi$ potentials calculated for
$^4$He and $^{208}$Pb nuclei in Fig.~\ref{fig:poto-nuclei}
for the two values of the cutoff mass, $1500$ MeV and $2000$ MeV.
The maximum depth of the $J/\Psi$ potential ranges roughly
from $18$ MeV to $24$ MeV.

Using the $J/\Psi$ potentials in nuclei obtained in this manner,
we next calculate the $J/\Psi$-nuclear bound state energies.
We follow the procedure applied in the earlier work~\cite{etaomega}
for the $\eta$- and $\omega$-nuclear bound states.
In this study, we focus entirely on the situation where
the $J/\Psi$ meson is produced nearly at rest (recoilless kinematics in experiments).
Then, it should be a very good approximation
to neglect the possible energy difference between the longitudinal and transverse
components~\cite{saitomega} of the $J/\Psi$ wave function ($\phi^\mu_{\Psi}$)
as was assumed for the $\omega$ meson case~\cite{saitomega}.
After imposing the Lorentz condition, $\partial_\mu \phi^\mu_{\Psi} = 0$,
to solve the Proca equation, aside from a possible width,
becomes equivalent to solving the Klein-Gordon equation with the reduced mass of
the system $\mu$ in vacuum~\cite{landau,Tabakin}:
\bge
\left[ \nabla^2 + E^2_\Psi - \mu^2 - 2\mu (m^*_\Psi(r) - m_\Psi) \right]\,
\phi_\Psi(\r) = 0,
\label{kgequation}
\ene
where $E_\Psi$ is the total energy of the $J/\Psi$ meson, and
$\mu = m_\Psi M/ (m_\Psi + M)$ with $m_\Psi$ ($M$) being the vacuum mass
of the $J/\Psi$ meson (nucleus).
Since in free space the width of $J/\Psi$ meson is $\sim 93$ keV~\cite{PDG},
we can ignore this tiny natural width in the following.
When the $J/\Psi$ meson is produced nearly at rest,
its dissociation process via $J/\Psi + N \to \Lambda_c^+ + \bar{D}$
is forbidden in the nucleus (the threshold energy
in free space is about $115$ MeV above). This is because the same number of
light quarks three participates in the initial and final states and hence,
the effects of the partial restoration of chiral symmetry
which reduces mostly the amount of the light quark condensates,
would affect a similar total mass reduction for
the initial and final states~\cite{Tsushima:1998ru,qmchadronmass}. Thus, the relative energy
($\sim$ 115 MeV) to the threshold would not be modified significantly.

We also note that once the $J/\Psi$ meson is bound in the nucleus,
the total energy of the system is below threshold for
nucleon knock-out and the whole system is stable.
The exception to this is the process $J/\Psi + N \to \eta_c + N$
which is exothermic.
Ref.~\cite{AlexPsiN} has provided an estimate of this cross section
from which we deduce a width for the bound $J/\Psi$ of order
$0.8$ MeV (nuclear matter at $\rho_B = \rho_0/2$).
We therefore expect that the bound $J/\Psi$, while not being
completely stable under the strong interaction, should be
narrow enough to be clearly observed. It would be worth while
to investigate this further.
Then, provided that experiments can be performed
to produce the $J/\Psi$ meson in recoilless kinematics,
the $J/\Psi$ meson is expected to be captured by
the nucleus into one of the bound states, which has
no strong interaction originated width.
This is a completely different and advantageous situation
compared to that of the $\eta$ and $\omega$ meson cases~\cite{etaomega}.
Thus, in the end, we may simply solve the Klein-Gordon equation with the reduced mass
Eq.~(\ref{kgequation}), without worrying about the width,
under the situation we consider now.
The calculation is carried out in momentum
space by the method developed in Ref.~\cite{landau,Tabakin}.
Bound state energies for the $J/\Psi$
meson, obtained solving the Klein-Gordon equation
Eq.~(\ref{kgequation}), are listed in Table~\ref{psienergy}.

\begin{table}[htbp]
\begin{center}
\caption{
Bound state energies calculated for the $J/\Psi$ meson,
solving the Klein-Gordon equation with the reduced mass
Eq.~(\ref{kgequation}).
Under the situation that the $J/\Psi$ meson is produced
in recoilless kinematics, the width due to the strong
interactions is all set to zero,
as well as its tiny natural width of $\sim 93$ keV~\cite{PDG}
in free space. (See also the text.)
}
\label{psienergy}

\begin{tabular}[t]{lc|c||c}
\hline \hline
 & &$\Lambda_{D,D^*}=1500$ MeV &$\Lambda_{D,D^*}=2000$ MeV\\
\hline \hline
 & &$E$ (MeV) &$E$ (MeV)\\
\hline

$^{4}_\Psi$He &1s &-4.19 &-5.74\\
\hline

$^{12}_\Psi$C &1s &-9.33 &-11.21\\
              &1p &-2.58 &-3.94\\
\hline
$^{16}_\Psi$O &1s &-11.23 &-13.26\\
              &1p &-5.11 &-6.81\\
\hline
$^{40}_\Psi$Ca &1s &-14.96 &-17.24\\
               &1p &-10.81 &-12.92\\
               &1d &-6.29  &-8.21 \\
               &2s &-5.63  &-7.48 \\
\hline
$^{90}_\Psi$Zr &1s &-16.38 &-18.69\\
               &1p &-13.84 &-16.07\\
               &1d &-10.92 &-13.06\\
               &2s &-10.11 &-12.22\\
\hline
$^{208}_\Psi$Pb &1s &-16.83 &-19.10\\
                &1p &-15.36 &-17.59\\
                &1d &-13.61 &-15.81\\
                &2s &-13.07 &-15.26\\
\hline \hline
\end{tabular}
\end{center}
\end{table}
%

The results in Table~\ref{psienergy} show that
the $J/\Psi$ are expected to form nuclear bound
states in all the nuclei considered.
This is insensitive to the values of cutoff mass used in the
form factors. It is very interesting to note that one can expect to find
a $J/\Psi$-$^4$He bound state. It will be possible
to search for this state as well
as the states in a $^{208}$Pb nucleus at JLab after the 12 GeV upgrade.
In addition, one can expect quite rich spectra
for medium and heavy mass nuclei.
Of course, the main issue is to produce the
$J/\Psi$ meson with nearly stopped kinematics, or
nearly zero momentum relative to the nucleus.
Since the present results imply that many nuclei should
form $J/\Psi$-nuclear bound states,
it may be possible to find such kinematics by careful selection
of the beam and target nuclei.

\section{Summary and discussion}
\label{sec_summary}

In this study, we have calculated the $J/\Psi$-nuclear potentials using
a local density approximation, with the inclusion of
the $DD$, $DD^*$ and $D^*D^*$ meson loops in the $J/\Psi$ self-energy.
The nuclear density distributions, as well as the in-medium $D$ and $D^*$
meson masses, are consistently obtained employing the quark-meson coupling model.
In the model, all the coupling constants between the applied meson mean fields and
the light quarks in the nucleon, as well as those in the $D$ and $D^*$ mesons,
are all equal and calibrated by the nuclear matter saturation properties.
Using the $J/\Psi$-nuclear potentials obtained in this manner,
we have solved the Klein-Gordon equation which is reduced
from the Proca equation, and obtained the $J/\Psi$-nuclear bound state energies.
For this, we have been able to set the strong interaction width of the
$J/\Psi$ meson to be zero (and neglect its tiny natural width of
$\sim 93$ keV in free space).
Combined with the generally advocated color-octet gluon-based attraction,
or QCD color van der Waals forces, we expect that the $J/\Psi$ meson will
form nuclear bound states and that the signal for the formation should be
experimentally very clear, provided that the $J/\Psi$ meson is produced in recoilless
kinematics.

\begin{acknowledgements}
The authors would like to thank S.~Brodsky and A.~Gal for helpful communications.
KT would like to acknowledge IFT UNESP (Brazil), where a part of this work
was carried out. The work of DHL was supported by Science Foundation of
Chinese University, while,
the work of GK was partially financed by CNPq and FAPESP (Brazilian agencies).
This work was also supported by the University of Adelaide and by
Australian Research Council through the
award of an Australian Laureate Fellowship (AWT).
\end{acknowledgements}


\end{document}